\documentclass[twocolumn,A4]{article}
\usepackage[dvips]{graphics}

\usepackage{setspace}
\usepackage{color}
\definecolor{gold}{rgb}{0.85,0.66,0}
\definecolor{dblue}{rgb}{0,0,0.8}

\begin{document}

\onecolumn

\begin{center}
{\bf{\Large {\textcolor{gold}{Quantum Transport in an Array of Mesoscopic 
Rings: Effect of Interface Geometry}}}}\\
~\\
{\textcolor{dblue}{Paramita Dutta}}$^{\dag}$, {\textcolor{dblue}{Santanu 
K. Maiti}}$^{\dag,\ddag,}$\footnote{{\bf Corresponding Author}: Santanu K. 
Maiti \\
$~$\hspace {0.45cm} Electronic mail: santanu.maiti@saha.ac.in}
{\textcolor{dblue}{and S. N. Karmakar}}$^{\dag}$ \\
~\\
{\em $^{\dag}$Theoretical Condensed Matter Physics Division,
Saha Institute of Nuclear Physics, \\
1/AF, Bidhannagar, Kolkata-700 064, India \\
$^{\ddag}$Department of Physics, Narasinha Dutt College,
129 Belilious Road, Howrah-711 101, India} \\
~\\
{\bf Abstract}
\end{center}
Electron transport properties are investigated in an array of mesoscopic 
rings, where each ring is threaded by a magnetic flux $\phi$. The array 
is attached to two semi-infinite one-dimensional metallic electrodes, 
namely, source and drain, where the rings are considered either in series 
or in parallel configuration. A simple tight-binding model is used to 
describe the system and all the calculations are done based on the Green's 
function formalism. Here, we present conductance-energy and current-voltage 
characteristics in terms of ring-to-electrode coupling strength, 
ring-electrode interface geometry and magnetic flux. Most interestingly 
it is observed that, typical current amplitude in an array of mesoscopic 
rings in the series configuration is much larger compared to that in  
parallel configuration of those rings. This feature is completely 
different from the classical analogy which may provide an important 
signature in designing nano-scale electronic devices.

\vskip 1cm
\begin{flushleft}
\end{flushleft}

\newpage
\twocolumn

\section{Introduction}

Study of quantum transport in low-dimensional systems has begun to 
flourish during the past few decades. All the basic features of 
electron transport in such systems solely depend on the concept of 
quantum interference effect, and it is generally preserved throughout 
the sample only for much smaller sizes, while the effect disappears 
for larger systems~\cite{imry1,imry2}. A normal metal mesoscopic ring 
is a very nice 
example where the electronic motion is confined and the transport 
becomes predominantly coherent~\cite{xia,buttiker,maiti1}. Several 
exotic phenomena are observed in such a ring system~\cite{peeters3,
peeters4,nitta}, especially in presence of magnetic flux, 
due to the effect of quantum interferences. In a very recent paper,
Peeters {\em et al.}~\cite{peeters2} have revealed this fact by doing 
a nice work on an array of mesoscopic rings. At thermodynamic 
equilibrium, electrons in a mesoscopic ring, threaded by magnetic flux 
$\phi$, can support non-decaying current even at non-zero temperature. 
This is the well-known phenomenon and the so-called persistent 
current~\cite{montam1,montam2,maiti2,peeters1,mailly,jari,levi} in 
normal metal ring. On the other hand, current trend of the 
miniaturization of electronic devices has resulted in intensive interest 
in characterization of these types of nanostructures. There are several 
methods for preparation of such rings. For example, gold rings can be 
designed by using templates of suitable structure in combination with 
metal deposition via ion beam etching~\cite{hobb,pearson}. More recently, 
Yan {\it et al.} have prepared gold rings by selective wetting of porous 
templates using polymer membranes~\cite{yan}. Though a few research groups 
have considered the serially~\cite{arunava,cui} or parallely connected ring 
systems~\cite{Liu} as their models of choice to study electron transport, 
yet several unexplained features are there in such systems. This motivates 
us to study electron transport in an array of mesoscopic rings, where each 
ring is threaded by an Aharonov-Bohm (AB) flux $\phi$. The electron 
transport properties through a bridge system was first studied 
theoretically in 1974~\cite{aviram}. Since then investigation on such 
two terminal devices is still going on and it is a major challenge in 
nano-electronics research.  

In the present paper we address the electronic transport properties in 
an array of mesoscopic rings which is attached to two semi-infinite 
one-dimensional ($1$D) metallic electrodes. To reveal the quantum 
interference effect on electron transport both the series and parallel
configurations are considered (for illustrative purposes see 
Figs.~\ref{series} and \ref{parallel}). An analytic approach based on 
a simple tight-binding model is given and all the calculations are done 
using the Green's function formalism~\cite{oreg1,oreg2,meir1,meir2,
baer1,baer2,baer3,maiti3,maiti4}. Here, we explore the conductance-energy
and current-voltage characteristics as functions of ring-electrode
interface geometry, coupling strength of the ring to the side attached
electrodes and magnetic flux $\phi$. Very interestingly we observe that,
typical current amplitude in an array of mesoscopic rings connected in
the series configuration is much higher than the array of such rings
connected in the parallel configuration. This behavior is completely 
different from the conventional resistors where the current through the
resistors in series configuration is much smaller than their parallel
arrangement. Our numerical study provides several key features which 
may be useful in designing tailor made nano-scale electronic devices.

We organize the paper specifically as follows. With a brief introduction
(Section $1$), in Section $2$, we describe the model and the theoretical
formulation for our calculations. Section $3$ presents the significant
results, and finally, we conclude our results in Section $4$.

\section{Model and synopsis of the theoretical background}

\begin{figure}[ht]
{\centering \resizebox*{7.5cm}{1.6cm}{\includegraphics{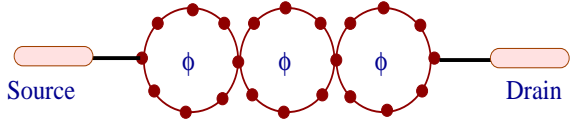}}\par}
\caption{(Color online). Schematic diagram of an array of symmetrically
connected mesoscopic rings (upper and lower arms of each ring contain
identical number of equally spaced lattice sites) in series configuration, 
where each ring is threaded by an AB flux $\phi$. The array is attached 
to two semi-infinite $1$D metallic electrodes, viz, source and drain. 
The filled red circles represent the position of the atomic sites.}
\label{series}
\end{figure}

\begin{figure}[ht]
{\centering \resizebox*{7.5cm}{4.2cm}{\includegraphics{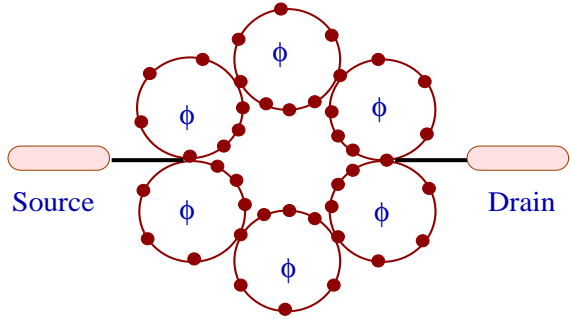}}\par}
\caption{(Color online). Schematic view of an array of symmetrically
connected mesoscopic rings (upper and lower arms of each ring have 
identical number of equally spaced lattice sites) in parallel 
configuration, where individual rings are subject to an AB flux $\phi$. 
The array is directly coupled to two electrodes, namely, source and drain. 
The filled red circles indicate the location of the atomic sites.}
\label{parallel}
\end{figure}

This section follows the theoretical formulation to study electron
transport in an array of mesoscopic rings. The rings are coupled
to each other through a single lattice site in such a way that
both the upper and lower arms of each ring contains equal number
of lattice sites. This configuration is the so-called symmetric
configuration. The rings are arranged in two different ways, series 
and parallel, those are schematically presented in Figs.~\ref{series} 
and \ref{parallel}, respectively. Each ring in the array is subjected 
to an AB flux $\phi$ (measured in unit of $\phi_0$ ($=ch/e$), the 
elementary flux-quantum) and the array is coupled to two semi-infinite 
$1$D metallic electrodes, namely, source and drain. 

To find the conductance ($g$) of the system, we use the Landauer 
conductance formula, where $g$ is expressed in terms of the 
transmission probability ($T$) of an electron as~\cite{datta2},
\begin{eqnarray}
g = \frac{2 e^2} {h} T 
\label{equ1}
\end{eqnarray}
This relation is valid only for low temperatures and bias voltages. In 
terms of the Green's function of the array, transmission probability 
can be expressed like~\cite{datta2,datta1},
\begin{eqnarray}
T = {\mbox{Tr}} [ \Gamma_S G^r_A \Gamma_D G^a_A ]
\label{transmission}
\end{eqnarray}
where, $G^r_A$ and $G^a_A$ are the retarded and advanced Green's functions 
of the array, respectively. Here, $\Gamma_S$ and $\Gamma_D$ are the
broadening matrices representing the couplings of the array to the
source and drain, respectively.

Now, the Green's function for the whole system i.e., array and the 
electrodes, is defined as,
\begin{eqnarray} 
G=\left(E - H \right)^{-1}
\end{eqnarray}
where, $E$ is the energy of the source electrons and $H$ is the 
Hamiltonian of the entire system which is of infinite dimension. 
To evaluate this Green's function, we require to calculate the inverse 
of an infinite dimensional matrix corresponding to the system consisting 
of a finite size array and the two semi-infinite electrodes, which is really
a difficult task. The full Hamiltonian can be partitioned into sub-matrices 
corresponding to the individual parts of the system like,
\begin{eqnarray}
H & = & H_A + H_S + H_D + \left(H_{SA}+H_{SA}^{\dag}\right) \nonumber \\
&  & + \left(H_{AD}+H_{AD}^{\dag}\right)
\end{eqnarray}
where, $H_A$, $H_S$ and $H_D$ correspond to the Hamiltonians of the 
array, source and drain, respectively. $H_{SA}$ and $H_{AD}$ represent 
source-to-array and array-to-drain coupling, respectively. Within the 
framework of non-interacting electron picture, the tight-binding 
Hamiltonian of the array can be expressed in the form,
\begin{eqnarray}
H_A = \sum_i \epsilon_i c_i^{\dag} c_i + \sum_{<ij>} t
\left(e^{i \theta} c_i^{\dag} c_j+ e^{-i\theta} c_j^{\dag} c_i \right)
\label{hamil}
\end{eqnarray}
where, $\theta=2 \pi \phi /N$ is the phase factor due the flux $\phi$
threaded by individual rings and $N$ is the total number of sites in 
each ring. Here, $t$ is the nearest-neighbor hopping integral, $\epsilon_i$ 
is the on-site energy and $c_i^{\dag} (c_i)$ is the creation (annihilation) 
operator of an electron at the site $i$. A similar kind of tight-binding 
Hamiltonian is also used, except the phase factor $\theta$, to illustrate 
the side attached $1$D perfect electrodes where the Hamiltonian is 
parametrized by constant on-site potential energy $\epsilon^{\prime}$ 
and nearest-neighbor hopping integral $t^{\prime}$. Like the Hamiltonian, 
Green's function can also be partitioned into sub-matrices and the 
effective Green's function for the array is written as~\cite{datta2,datta1},
\begin{eqnarray}
G_A=\left(E -H_A-\Sigma_S - \Sigma_D \right)^{-1}
\label{greenarray}
\end{eqnarray}
where, $\Sigma_S$ and $\Sigma_D$ are the self-energies due to the coupling
of the rings to the source and drain, respectively. 

The broadening matrices $\Gamma_S(E)$ and $\Gamma_D(E)$, in 
Eq.~\ref{transmission}, are defined as the difference between the retarded 
and advanced self-energies of the electrodes. Explicitly, one can write,
\begin{eqnarray}
\Gamma_{S(D)}(E) = i \left[\Sigma^r_{S(D)}(E)-\Sigma^a_{S(D)}(E)\right]
\label{gamma}
\end{eqnarray}
and the self-energies can be expressed as~\cite{datta2},
\begin{eqnarray}
\Sigma^r_{S(D)} = \Lambda_{S(D)} - i \Delta_{S(D)}
\label{sigma}
\end{eqnarray}
where, the real parts ($\Lambda_{S(D)}$) correspond to the shift of the 
energy levels of the array and the imaginary parts ($\Delta_{S(D)}$)
represent the broadening of these energy levels. We do our calculations 
only at absolute zero temperature. But all these results are also valid 
at low temperatures for which $k_B T$ is much smaller than the average 
spacing of the energy levels. Using Eq.~\ref{sigma}, the broadening 
matrices can now be written as,
\begin{eqnarray}
\Gamma_{S(D)} = -2 \mbox{Im} \left(\Sigma^r_{S(D)}\right)
\label{gammasd}
\end{eqnarray}
To evaluate the current passing through the array, we use the
relation~\cite{datta2},
\begin{eqnarray}
I(V) = \frac{2e}{h} \int \limits_{-\infty}^{\infty} 
\left(f_S-f_D\right) T(E)~ dE
\label{equ10}
\end{eqnarray}
where, $f_{S(D)}=f\left(E-\mu_{S(D)}\right)$ gives the Fermi 
distribution function with the electrochemical potential 
$\mu_{S(D)}=E_F\pm eV/2$. $V$ is the applied bias voltage. Here, we 
assume that the entire voltage is dropped across the conductor-electrode 
interfaces, and, this assumption does not affect significantly the 
current-voltage characteristics in such a small array~\cite{tian}. 
Throughout the calculation, we choose the units where $c=e=h=1$ and 
set the Fermi energy $E_F$ at $0$. 

\section{Numerical results and discussion}

\subsection{Conductance-energy characteristics}

In order to illustrate the results, we begin our discussion by 
mentioning the values of the different parameters used for the 
numerical calculations.
In the ring, the site energy $\epsilon_i$ is fixed to $0$ and the 
nearest-neighbor hopping integral $t$ is set to $3$. While, for the 
side-attached electrodes the on-site energy ($\epsilon^{\prime}$) and 
the nearest-neighbor hopping strength ($t^{\prime}$) are chosen as 
$0$ and $4$, respectively. Throughout our study, we narrate all the 
essential features of electron transport in the two different regimes 
depending on the strength of the coupling of the ring to the source 
and drain. One is the weak-coupling regime defined by the condition 
$\tau_{S,D}<<t$. For this regime we choose $\tau_S=\tau_D=0.5$. Other 
one is the strong-coupling regime defined by the relation $\tau_{S,D} 
\sim t$. In this particular limit, we set the values of the parameters 
as $\tau_S=\tau_D=2.5$. Here, $\tau_S$ and $\tau_D$ describe the 
coupling strengths of the ring to the source and drain, respectively.
In addition to these, we also introduce two other parameters $N$ and
$M$ to describe the size of the array, where they correspond to the
ring size and total number of rings in the array, respectively.

To explore the basic mechanisms of electron transport in an array of
\begin{figure}[ht]
{\centering \resizebox*{7.8cm}{9.5cm}{\includegraphics{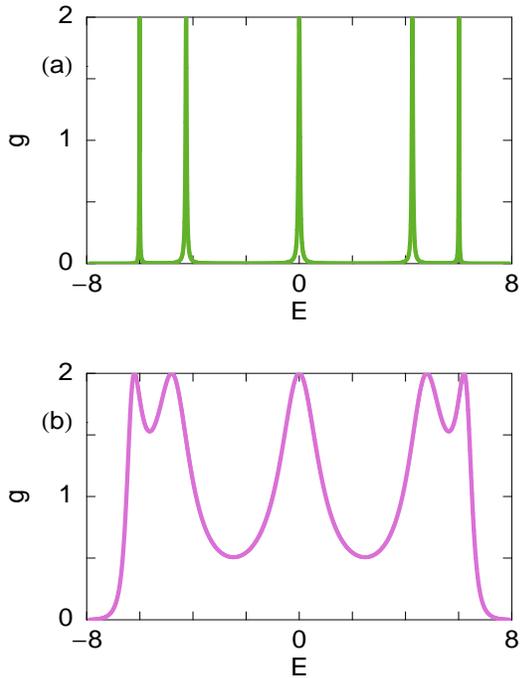}}\par}
\caption{(Color online). $g$-$E$ spectra for a symmetrically connected 
mesoscopic ring with $\phi=0$, where we set $M=1$ and $N=8$. (a) and (b) 
correspond to the weak- and strong-coupling limits, respectively.}
\label{onering}
\end{figure}
mesoscopic rings, we start with the results for a single ring ($M=1$) 
and then two rings ($M=2$) which are directly coupled to each other.

As illustrative examples, in Fig.~\ref{onering} we plot the conductance 
$g$ as a function of injecting electron energy $E$ for a symmetrically 
connected mesoscopic ring, where (a) and (b) correspond to the results 
for the weak- and strong-coupling limits, respectively. Here we fix 
$N=8$ (total number of atomic sites in the ring, where the upper and 
lower arms contain $3$ atomic sites). All these results are calculated 
in the absence of AB flux $\phi$. From the $g$-$E$ spectra it is observed 
that, in the case of weak-coupling limit, conductance exhibits sharp 
resonant peaks for some particular energies, while it ($g$) disappears 
for the other energies. Thus, a fine tuning in the energy scale is 
required to get electron conduction across the ring. At resonances, 
conductance reaches the value $2$, and accordingly, the transmission 
\begin{figure}[ht]
{\centering \resizebox*{7.8cm}{9.5cm}{\includegraphics{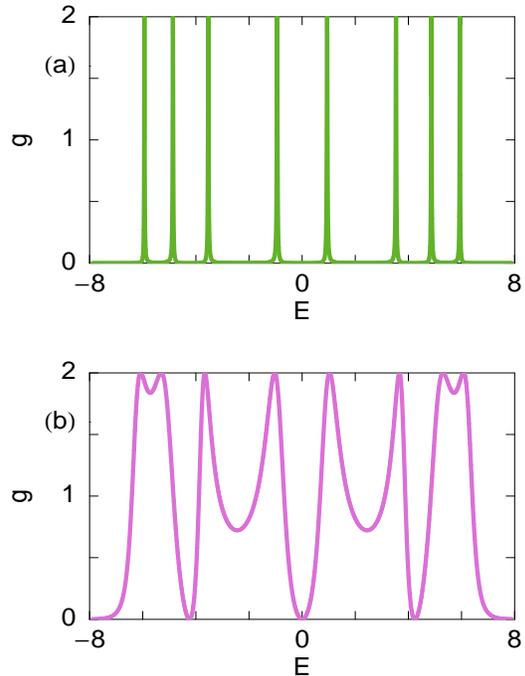}}\par}
\caption{(Color online). $g$-$E$ spectra for a symmetrically connected 
mesoscopic ring with $\phi=0.2$, where we set $M=1$ and $N=8$. (a) and (b) 
correspond to the weak- and strong-coupling limits, respectively.}
\label{oneringphi}
\end{figure}
probability $T$ approaches to unity, since the equation $g=2T$ is 
satisfied from the Landauer conductance formula (see Eq.~\ref{equ1} 
with $e=h=1$ in our present formulation). All these resonant peaks are 
associated with the energy eigenvalues of the ring, and therefore, we 
can predict that the conductance spectrum manifests itself the 
fingerprint of the electronic structure of the ring. The situation 
becomes quite interesting as long as the coupling strength of the ring 
to the side attached electrodes is increased. In the limit of 
strong-coupling, all the resonant peaks get substantial widths compared 
to the weak-coupling case. The contribution to the broadening of the 
resonant peaks in this strong-coupling limit comes from the imaginary 
parts of the self-energies $\Sigma_S$ and $\Sigma_D$~\cite{datta2}. 
Hence, by tuning the coupling strength, we can get the electron 
transmission across the ring for the wider range of energies which 
provides an important signature in the study of current-voltage 
($I$-$V$) characteristics.

To emphasize the effect of AB flux on electron transport through such
a ring, now we describe the results presented in Fig.~\ref{oneringphi}.
The results are computed for the same ring i.e., $N=8$, considering 
$\phi=0.2$, where (a) and (b) correspond to the weak- and strong-coupling 
cases, respectively. The effects of coupling on the resonant widths 
become exactly identical as in Fig.~\ref{onering}. But, quite 
interestingly, we observe that a global gap in the conductance
spectrum appears across the energy $E=0$ as long as the flux $\phi$ is
applied in the ring. This feature is clearly noticed by comparing 
the results plotted in Figs.~\ref{onering} and \ref{oneringphi}. It is 
also verified that the gap across the energy $E=0$ increases gradually 
with the increase in $\phi$, and at the typical flux $\phi=\phi_0/2$, 
conductance exactly vanishes for the entire energy range~\cite{maiti3,
maiti4}. This vanishing behavior of the transmission probability at 
$\phi=\phi_0/2$ for a symmetrically connected ring, where the upper 
and lower arms are identical in nature, can be obtained very easily 
by a simple mathematical calculation as follows. For a symmetrically 
connected ring, the wave functions passing through the upper and lower 
arms of the ring are given by,
\begin{eqnarray}
\psi_1 & = & \psi_0 e^{\frac{ie}{\hbar c} \int \limits_{\gamma_1} 
\vec{A}.\vec{dr}} \nonumber \\
\psi_2 & = & \psi_0 e^{\frac{ie}{\hbar c} \int \limits_{\gamma_2} 
\vec{A}.\vec{dr}} 
\label{equ11}
\end{eqnarray}
where, $\gamma_1$ and $\gamma_2$ are used to indicate the two different
paths of electron propagation along the two arms of the ring. $\psi_0$
denotes the wave function in absence of magnetic flux $\phi$ and it is
same for both upper and lower arms as the ring is symmetrically coupled
to the electrodes. $\vec{A}$ is the vector potential associated with the
\begin{figure}[ht]
{\centering \resizebox*{7.8cm}{9.5cm}{\includegraphics{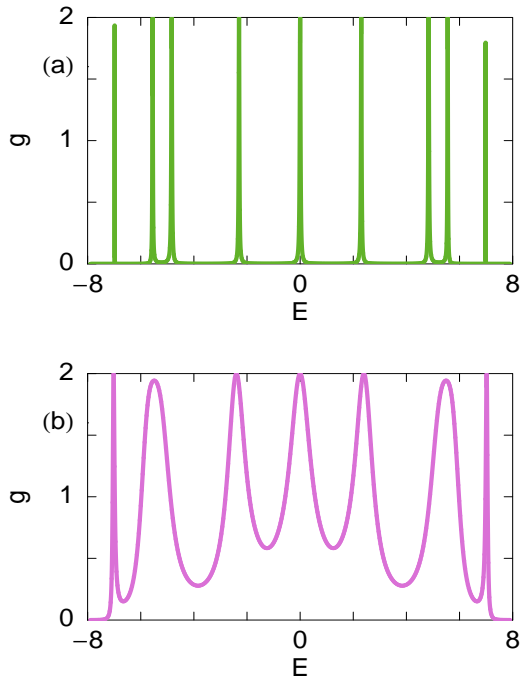}}\par}
\caption{(Color online). $g$-$E$ spectra for an array of symmetrically
connected mesoscopic rings with $\phi=0$ in the series configuration,
where we set $M=2$ and $N=8$. (a) and (b) correspond to the weak- and
strong-coupling limits, respectively.}
\label{tworing}
\end{figure}
magnetic field $\vec{B}$ by the relation $\vec{B}= \vec{\nabla} \times 
\vec{A}$. Hence the probability amplitude of finding the electron passing
through the ring can be calculated as,
\begin{equation}
|\psi_1 + \psi_2|^2 = 2|\psi_0|^2 + 2|\psi_0|^2 \cos \left({\frac{2\pi 
\phi}{\phi_0}}\right)
\label{equ12}
\end{equation}
where, $\phi = \oint \vec{A}.\vec{dr} = \int \int \vec{B}.\vec{ds}$
is the flux enclosed by the ring. From Eq.~\ref{equ12} it is clearly 
observed that at $\phi=\phi_0/2$ the transmission probability of an 
electron through a symmetrically connected ring drops exactly to zero.

\begin{figure}[ht]
{\centering \resizebox*{7.8cm}{9.5cm}{\includegraphics{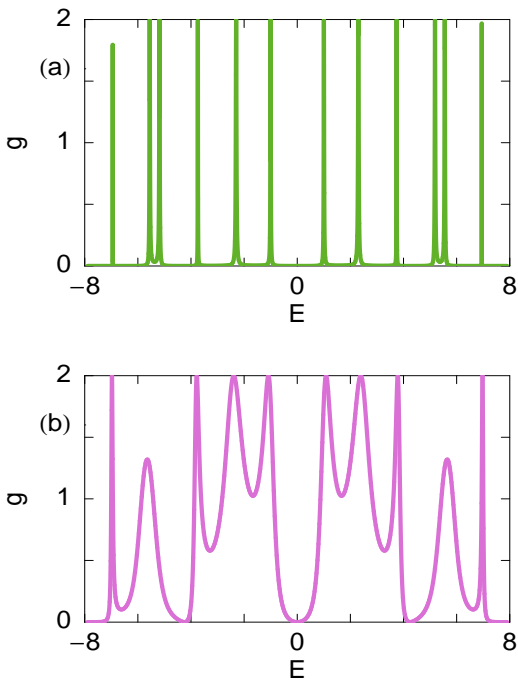}}\par}
\caption{(Color online). $g$-$E$ spectra for an array of symmetrically
connected mesoscopic rings with $\phi=0.2$ in the series configuration,
where we set $M=2$ and $N=8$. (a) and (b) correspond to the weak- and
strong-coupling limits, respectively.}
\label{tworingphi}
\end{figure}

Now we study the conductance-energy characteristics for an array of
two rings ($M=2$) having $3$ sites in each arm of the individual rings. 
In Fig.~\ref{tworing} we show the results for $\phi=0$, while for 
$\phi=0.2$ the results are given in Fig.~\ref{tworingphi}. Both for
the weak- and strong-coupling limits we get exactly similar features
of resonant peaks i.e., sharp in weak-coupling regime and broadened
in strong-coupling regime, as presented in Figs.~\ref{onering} and
\ref{oneringphi}. Also, a finite gap across the energy $E=0$ appears
for any non-zero value of $\phi$ (for instance see Fig.~\ref{tworingphi}).
The only difference is that for such a two ring system number of resonant
peaks, associated with the energy eigenvalues, in the conductance 
spectra gets increased compared to the one ring system. Apart from these 
features, from Figs.~\ref{tworing} and \ref{tworingphi} we see that for 
some specific energy values resonances are of Fano type, where a sharp 
peak followed by a sharp deep is observed. With the increase of 
ring-to-electrode coupling strength, positions of these peaks and deeps 
and also their heights are unchanged. Only the widths get broadened with 
the increase of coupling strength. This behavior is quite different from 
the other resonant peaks where a sharp peak is not followed by a sharp deep.

Following the above conductance-energy characteristics of one and two
\begin{figure}[ht]
{\centering \resizebox*{7.8cm}{9.5cm}{\includegraphics{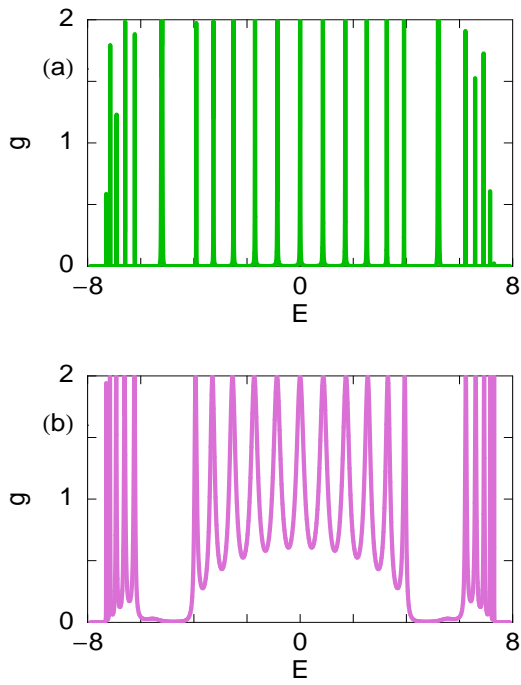}}\par}
\caption{(Color online). $g$-$E$ spectra for an array of symmetrically
connected mesoscopic rings with $\phi=0$ in the series configuration,
where we set $M=6$ and $N=8$. (a) and (b) correspond to the weak- and
strong-coupling limits, respectively.}
\label{ser}
\end{figure}
ring systems, now we focus our study on larger ring systems where
the rings are arrayed both in series and parallel configurations
to illustrate the effect of quantum interference on electron transport.

\begin{figure}[ht]
{\centering \resizebox*{7.8cm}{9.5cm}{\includegraphics{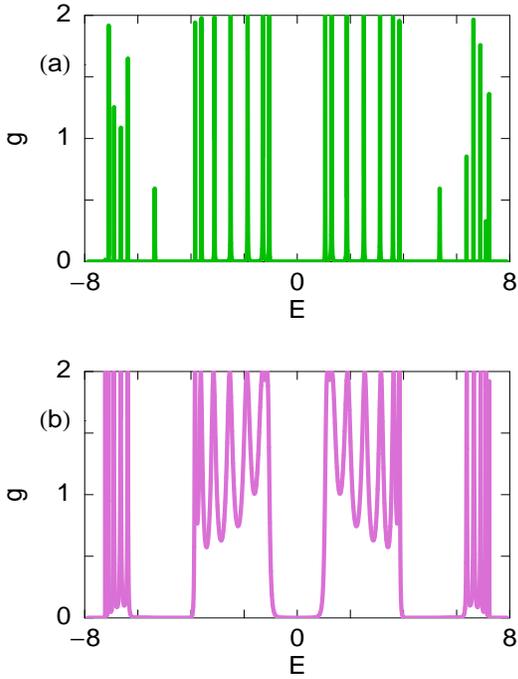}}\par}
\caption{(Color online). $g$-$E$ spectra for an array of symmetrically
connected mesoscopic rings with $\phi=0.2$ in the series configuration,
where we set $M=6$ and $N=8$. (a) and (b) correspond to the weak- and
strong-coupling limits, respectively.}
\label{serphi}
\end{figure}

\begin{figure}[ht]
{\centering \resizebox*{7.8cm}{9.5cm}{\includegraphics{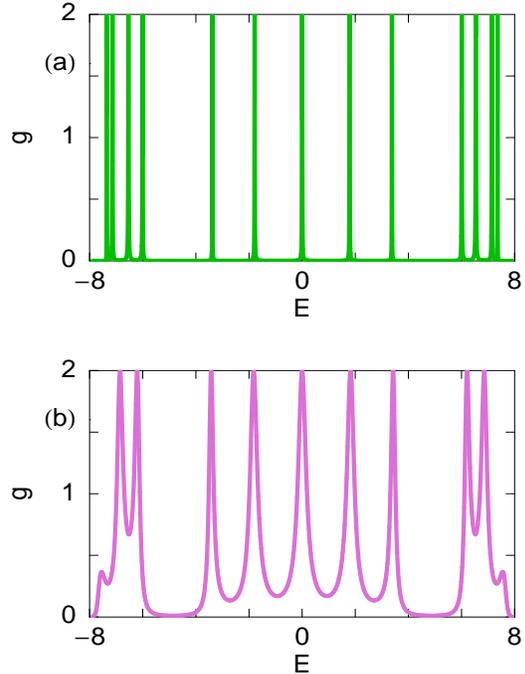}}\par}
\caption{(Color online). $g$-$E$ characteristics for an array of
symmetrically connected mesoscopic rings with $\phi=0$ in the parallel
configuration, where (a) and (b) correspond to the weak- and
strong-coupling limits, respectively. All the other parameters are same
as in Fig.~\ref{ser}.}
\label{par}
\end{figure}

\begin{figure}[ht]
{\centering \resizebox*{7.8cm}{9.5cm}{\includegraphics{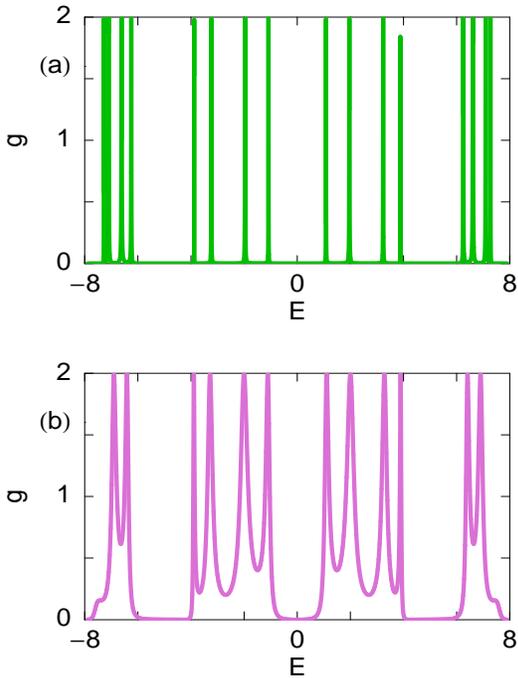}}\par}
\caption{(Color online). $g$-$E$ characteristics for an array of
symmetrically connected mesoscopic rings with $\phi=0.2$ in the parallel
configuration, where (a) and (b) correspond to the weak- and
strong-coupling limits, respectively. All the other parameters are same
as in Fig.~\ref{ser}.}
\label{parphi}
\end{figure}

As representative examples, in Fig.~\ref{ser} we plot the 
conductance-energy characteristics for an array of mesoscopic rings,
where the rings are arranged in series configuration. The results
are depicted for the array with $M=6$ and $N=8$ (upper and lower arms
of each ring have $3$ atomic sites) in the absence of $\phi$, where 
(a) and (b) correspond to the weak- and strong-coupling cases, 
respectively. For the same array, we display the results in 
Fig.~\ref{serphi}, considering $\phi=0.2$. On the other hand,
the variation of conductance $g$ as a function of energy $E$ in the
absence of $\phi$ for an array of rings, arranged in parallel 
configuration, is given in Fig.~\ref{par}, where (a) and (b) represent 
the identical meaning as in Fig.~\ref{ser}. Here we fix $M=6$ ($3$ rings 
in each branch) and $N=8$ (upper and lower arms of each ring contain 
$3$ atomic sites). Both for the two different coupling limits, the nature
of the resonant peaks is exactly identical as studied earlier
(Figs.~\ref{onering} and \ref{oneringphi}). Comparing the results
presented in Figs.~\ref{ser} and \ref{par}, we notice that in the 
parallel configuration of the rings, several resonant peaks disappear 
compared to their (rings) series configuration. This is solely due to 
the effect of quantum interference among the electronic waves passing 
through different arms of the rings and it provides an interesting feature
in electron transport which can be much more clearly explained from
our current-voltage characteristics (in sub-section $3.2$). In the
same footing, as above, in Fig.~\ref{parphi} we plot the $g$-$E$
characteristics for the same array ($M=6$ and $N=8$) in the presence
of $\phi$ ($\phi=0.2$), where different curves represent the identical
meaning as in Fig.~\ref{par}. Similar to the case of $\phi=0$, the 
number of resonant peaks in the parallel configuration also decreases 
compared to the series configuration (Fig.~\ref{serphi}) in the presence
of $\phi$. Thus, it is manifested that the number of resonant peaks in 
parallel configuration is always reduced than the series configuration 
irrespective of $\phi$ threaded by the rings. In Fig.~\ref{parallel}, 
the rings are arranged in parallel configuration where each ring is 
penetrated by an AB flux $\phi$. In such an arrangement a bigger loop 
is formed where we do not apply any magnetic flux for our present 
discussion. We can also apply a magnetic flux through this bigger 
loop and in that case an electron experiences two magnetic phases 
during the propagation through upper and lower arms of the rings. 
Even in that case i.e., in the presence of two fluxes, the number of 
resonant resonant peaks in parallel configuration is much smaller 
compared to the series one. Thus in short we can say that, appearance 
of more resonant peaks in series configuration than the parallel one 
is valid for the cases when (i) there is no flux through individual 
rings as well as in bigger loop, (ii) in presence of AB fluxes in 
the smaller rings only, but, not in the bigger ring and (iii) in 
presence of AB fluxes through identical rings as well as in bigger 
ring. All these features can also be justified through proper 
experimental arrangement.

\subsection{Current-voltage characteristics}

\begin{figure}[ht]
{\centering \resizebox*{7.8cm}{9.8cm}{\includegraphics{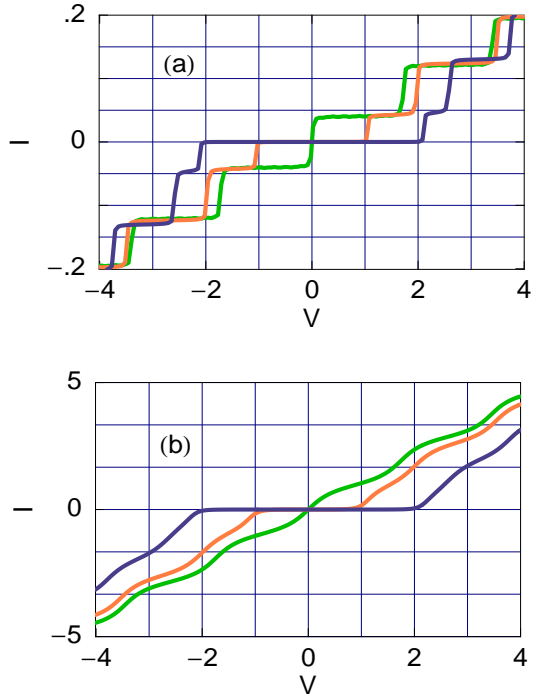}}\par}
\caption{(Color online). $I$-$V$ characteristics for an array of
symmetrically connected mesoscopic rings in the series configuration,
where we set $M=6$ and $N=8$. The green, orange and blue curves
correspond to $\phi=0$, $0.1$ and $0.2$, respectively. (a) weak-coupling
limit and (b) strong-coupling limit.}
\label{curr1}
\end{figure}

All the basic features of electron transport obtained from conductance 
versus energy spectra can be explained in a better way through the 
current-voltage ($I$-$V$) characteristics. The current across the array 
is determined by integrating over the transmission curve according to 
Eq.~\ref{equ10} which is not restricted in the linear response regime, 
but it is of great significance in determining the shape of the full 
current-voltage characteristics. As representative examples, in 
Fig.~\ref{curr1} we display
the current-voltage ($I$-$V$) characteristics for an array of symmetrically
connected mesoscopic rings in the series configuration, where (a) and (b)
correspond to the results for the weak- and strong ring-to-electrode
coupling limits, respectively. The currents are evaluated for an array
considering $M=6$ and $N=8$, where the green, orange and blue lines 
correspond to the currents for $\phi=0$, $0.1$ and $0.2$, respectively. 
From the results it is observed that, in the case of weak-coupling, 
current exhibits staircase-like structure with sharp steps as a function 
of the applied bias voltage. This is due to the presence of fine resonant 
peaks in the conductance spectra, as the current is obtained from 
integration method over transmission function $T$. With the increase 
\begin{figure}[ht]
{\centering \resizebox*{7.8cm}{9.8cm}{\includegraphics{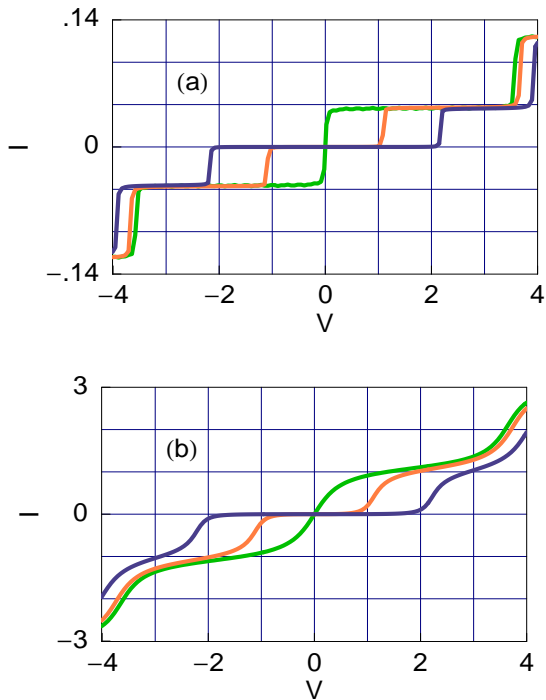}}\par}
\caption{(Color online). $I$-$V$ characteristics for an array of
symmetrically connected mesoscopic rings in the parallel configuration,
where we set $M=6$ and $N=8$. The curves in (a) and (b) correspond to
the identical meaning as in Fig.~\ref{curr1}.}
\label{curr2}
\end{figure}
in applied bias voltage $V$, the difference in chemical potentials of 
the two electrodes $(\mu_1-\mu_2)$ increases, allowing more number 
of energy levels to fall in that range, and therefore, more number of
energy channels are accessible to the injected electrons to pass 
through the array from the source to drain. Incorporation of a single
discrete energy level i.e., a discrete quantized conduction channel,
between the range $(\mu_1-\mu_2)$ provides a jump in the $I$-$V$
characteristics.
In addition to this step-like feature, we also observe that in the absence
of $\phi$, current shows non-zero value i.e., electron is allowed to pass 
through the array as long as the bias voltage is switched on (see the green 
curve). While, in the presence of $\phi$, the conduction of an electron
is started beyond some finite value of the bias voltage, the so-called
threshold voltage $V_{th}$ (see the orange and blue lines), which is 
clear from the $g$-$E$ spectrum shown in Fig.~\ref{serphi}(a). Thus, by
changing the value of $\phi$, $V_{th}$ can be regulated in a controlled
way. This behavior may be useful in designing nano-scale electronic
devices. The nature of the $I$-$V$ spectra changes significantly in the
case of strong-coupling (Fig.~\ref{curr1}(b)). Here, the current varies
almost continuously as a function of the bias voltage and achieves much
higher current amplitude, even an order of magnitude, compared to the 
weak-coupling case. Therefore, we can emphasize that for a fixed bias
voltage $V$, one can enhance the current amplitude through the bridge
system by tuning the ring-to-electrode coupling strength.

At the end, we describe the effect of interface geometry on the behavior 
of $I$-$V$ characteristics. The results are plotted in Fig.~\ref{curr2},
where (a) and (b) correspond to the identical meaning as in Fig.~\ref{curr1}. 
In this parallel configuration, both for the weak- and strong-coupling 
regimes, behavior of the current i.e., step-like and continuous nature,
is exactly similar to that as discussed in the case of series 
configuration. But, the key observation is that, for a fixed ring-electrode
coupling strength and applied bias voltage, the current amplitude obtained
in the parallel configuration is much smaller compared to that in the 
series configuration. This anomalous behavior is completely different 
from the traditional resistors where the current amplitude obtained in
the case of parallel arrangement is much higher than that in series 
configuration. All these features of electron transport may provide
some physical insight to study transport properties in array-like
geometries.

\section{Concluding remarks}

To conclude, we have addressed the electronic transport properties in 
an array of mesoscopic rings, where each ring is threaded by an AB 
flux $\phi$. Both the series and parallel configurations of the rings
are considered to reveal the quantum interference effect on electron 
transport. A parametric approach based on the tight-binding framework
is given where all the calculations are done through the Green's function
formalism. Our exact numerical calculations describe conductance-energy 
and current-voltage characteristics in aspects of (a) ring-to-electrode 
coupling strength, (b) ring-electrode interface geometry and (c) magnetic 
flux. Most significantly we predict that, in a series configuration of
mesoscopic rings the typical current amplitude is much higher compared 
to their parallel configuration. This phenomenon is completely different
from the traditional one. Here, we have presented our results only for 
the array of symmetrically connected mesoscopic rings. All these features
are also valid for an array where the rings are coupled asymmetrically
to each other and due to the obvious reason we do not plot those results 
further. Our exact analysis may provide some important signatures to study 
electron transport in nano-scale systems.

Throughout our work, we have explored the conductance-energy and
current-voltage characteristics for an array of $6$ mesoscopic rings
each having $8$ atomic sites. In our model calculations, these typical
numbers are chosen only for the sake of simplicity. Though the
results presented here change numerically with the ring size and total
number of rings, but all the basic features remain exactly invariant. 
To be more specific, it is important to note that, in real situation 
the experimentally achievable rings have typical diameters within the 
range $0.4$-$0.6$ $\mu$m. In such a small ring, a high magnetic field 
is required to produce a quantum flux. To overcome this situation, Hod 
{\em et al.} have studied extensively and proposed how to construct 
nanometer scale devices, based on Aharonov-Bohm interferometry, those 
can be operated in moderate magnetic fields~\cite{baer4,baer5,baer6,baer7}. 
In addition, it is also important to note that here we have done all 
the numerical calculations for some specific values of the parameters. 
One can also compute these results for some other typical values of the 
parameters. In that case only the numerical values will be changed, but 
the basic characteristics remain unaltered.

In the present paper we have done all the calculations by ignoring
the effects of the temperature, disorder, electron-electron correlation, 
etc. The effect of the temperature has already been reported earlier, 
and, it has been examined that the presented results will not change 
significantly even at finite (low) temperature, since the broadening 
of the energy levels of the ring system due to its coupling to the 
electrodes will be much larger than that of the thermal broadening.
In presence of disorder, scattering process appears in the arms of 
the rings that can influence the electronic phases, and accordingly,
the quantum interference effect is disturbed. All the above pictures 
are also valid if electron-electron interaction is taken into account. 
In presence of electronic correlation, the on-site Coulomb repulsive 
energy $U$ gives a renormalization of the site energies~\cite{datta3}. 
Depending on the strength of the nearest-neighbor hopping integral 
($t$) compared to the on-site Coulomb interaction ($U$) different 
regimes appear. For the case $t/U<<1$, the resonances and 
anti-resonances would split into two distinct narrow bands separated 
by the on-site Coulomb energy. On the other hand, for the case where 
$t/U>>1$, the resonances and anti-resonances would occur in pairs. 
In a recent work, Montambaux {\em et al.}~\cite{mont} have studied 
elaborately the effect of electron-electron correlation on electron 
transport for some arrays of connected mesoscopic metallic rings. In 
this work, they have studied how at low temperatures, decoherence is 
limited by electron-electron interaction. Finally, we would like to 
say that we need further study in such systems by incorporating all 
these effects.

\end{document}